# Recommendations for repositories and scientific gateways from a neuroscience perspective

Malin Sandström[1,a], Mathew Abrams[1,b], Jan Bjaalie[2,c], Mona Hicks[3,d], David Kennedy[4,e], Arvind Kumar[5,f], JB Poline[6,g], Prasun Roy[7,h], Paul Tiesinga[8,i], Thomas Wachtler[9,j], Wojtek Goscinski[10,11,k]

1. INCF Secretariat, Karolinska Institutet, Stockholm, Sweden
2. Institute of Basic Medical Sciences, University of Oslo, Oslo, Norway
3. One Mind, US
4. Department of Psychiatry, University of Massachusetts Chan Medical School, North. Worcester, USA
5. Div. of Computational Science and Technology, School of Electrical Engineering and Computer Science, KTH Royal Institute of Technology, Stockholm, Sweden
6. Montreal Neurological Institute, Faculty of Medicine and Health Sciences, McGill University, Montreal, Canada
7. Computational Neuroscience & Neuroimaging Laboratory, School of Bio-Medical Engineering, Indian Institute of Technology - I.I.T. (B.H.U.), Varanasi, India
8. Department of Neuroinformatics, Donders Centre for Neuroscience, Faculty of Science, Radboud University, Nijmegen, The Netherlands
9. Department of Biology II, Ludwig-Maximilians-Universität München, Planegg-Martinsried, Germany
10. National Imaging Facility, University of Queensland, St Lucia, Australia
11. Monash eResearch Centre, Monash University, Clayton, 3800, Australia

Author ORCIDs
a: ORCID:0000-0002-8464-2494
b: ORCID:0000-0001-9438-9691
c: ORCID:0000-0001-7899-906X
d: ORCID:0000-0001-7418-6221
e: ORCID:0000-0002-9377-0797
f: ORCID:0000-0002-8044-9195
g: ORCID:0000-0002-9794-749X
h: ORCID:0000-0002-9794-749X
i: ORCID:0000-0003-4509-6870
j: ORCID:0000-0003-2015-6590
k: ORCID:0000-0001-6587-1016

## Abstract

Digital services such as repositories and science gateways have become key resources for the neuroscience community, but users often have a hard time orienting themselves in the service landscape to find the best fit for their particular needs. INCF (International Neuroinformatics Coordinating Facility) has developed a set of recommendations and associated criteria for choosing or setting up and running a repository or scientific gateway, intended for the neuroscience community, with a FAIR neuroscience perspective. These recommendations have neurosciences as their primary use case but are often general. Considering the perspectives of researchers and providers of repositories as well as scientific gateways, the recommendations harmonize and complement existing work on criteria for repositories and best practices. The recommendations cover a range of important areas including accessibility, licensing, community responsibility and technical and financial sustainability of a service.

## Introduction

Digital data repositories play an important role in the archiving, management, analysis and sharing of research data. They provide stable, long-term storage, can improve data quality through active curation, can increase the discoverability and reusability of data through the use of controlled terms and standardized metadata, and make it easier to request and transfer data, as well as removing or lowering barriers to reuse and collaboration. Data shared in repositories is more often cited than data shared by other means, like supplements[1].

Modern neuroscience datasets are often in the gigabyte range and commonly reach the terabyte level[2,3], and in some cases the petabyte level[4]. That amount of data is hard to handle without accompanying computational power, data viewing and data analysis capabilities in the same place. We refer to enhanced repositories that provide such resources as scientific gateways, to distinguish them from regular repositories. Throughout, we will use the term "services" to refer jointly to repositories and scientific gateways. Scientific gateways offer computational resources and built-in software resources, sometimes also data visualization, custom data analysis and/or workflow composition. They usually have user accounts and might store data that is only available to logged-in users.

Besides hosting data and providing computational power, repositories and scientific gateways are also important for supporting research reproducibility and replicability; they can preserve data and computational research outcomes that might otherwise be lost or become unfindable over time, and make it realistically possible to redo the same analyses or computational experiments. Openly available data storage and computational resources also have the possibility to become a driver for increasing diversity and equality in science, as they help counteract differences in access to hardware, tools and resources.

The current repository landscape is quite diverse and varied, and the many different possible choices may thus confuse the intended users. Recommendations for community specific repositories vary from one body to another, and scientific gateways are not included at all in most recommendations. Researchers are often asked to pick the resource that fits them best, but feel they have little guidance to do so[5].

Therefore, INCF has developed selection criteria and associated recommendations for the neuroscience community, with a FAIR neuroscience perspective, and tried to harmonize them with existing work on criteria for repository selection and best practices from FAIRsharing[6], FORCE11[7] and Coalition of Open Access Repositories[8]. We are also taking into account the feedback received in our workshop "Towards neuroscience-centered selection criteria for data repositories and scientific gateways" held on April 26, 2021 at the yearly INCF Assembly[5].

The full set of criteria is available on the INCF portal (https://www.incf.org/criteria-checklist).

Our aim is two-fold: we want to help neuroscience researchers and students choose good services for their specific use cases; we also want to help service providers make good and future-proof decisions for setup and operations.

The full technical aspects of setting up and running a FAIR service is outside the scope of this comment; we recommend that service providers consult an external resource such as the FAIRSFAIR Basic Framework on FAIRness of services[9] or the COAR Community Framework for Good Practices in Repositories[8].

---

BOX 1: Recommendations

Ensure discoverability and transparency in ownership & service usage statistics

Clearly communicate access & reuse conditions

Consider ethical requirements for authorship transparency and sensitive data

Follow best practices for licensing & responsibility

Ensure accessibility and interoperability

Build capabilities for reproducibility, replicability, reuse

Excel in documentation & user support

Be transparent in governance

Involve community in governance & decision making

Be transparent on sustainability - financial & technical

---

# Ensure discoverability and transparency in ownership & service usage statistics

Finding suitable repositories or scientific gateways becomes much easier when each clearly states what services it contributes to the community, and identifies its target users. Are these services just offering data or also accepting submissions, what types of submissions are accepted, what additional services are offered? Which jurisdiction applies, and who supports the service?

We recommend that services provide a clear and concise description that outlines the resource features, identifies the intended community, and states who supports the service. We also recommend that service providers are transparent in communicating their usage data and usage data history, ideally in a machine readable way, so that historic and current usage can be easily understood and extracted. The level and pattern of usage are important proxies for harder to judge criteria such as community importance, community relevance and impact.

We also recommend services to register in relevant repository registries, such as [Re3data](#) or [FAIRSharing](#), and to consider participating in a certification for trustworthiness. The most common certification, Core Trust Seal, has requirements which align with a number of the FAIR requirements.[10]

# Clearly communicate conditions for access & reuse

Finding the best possible service for one's use case is of little worth if the service then cannot be used, because of geographical, financial, technical or administrative barriers. We recommend that service providers clearly state access and deposit conditions and the cost, if any, of usage. We also recommend that a service clearly and prominently communicates all of the file formats and metadata formats it accepts and uses.

To facilitate reuse, derived data and other downloaded resources should by default have clear standard licenses [see 'Follow best practices for licensing & responsibility'].

# Consider ethical requirements for authorship transparency and sensitive data

Authorship is a core component of any metadata set. Services should be able to handle authorship changes, especially for software where authorship may change for each update. For living objects that can be updated, data as well as software, we recommend that the service should make it possible to change authorship with any update, and be transparent with the change history of authorship. Ideally the service also will provide contact information for at least one author.

Proof of ethics approval should be required for sensitive data, and services that accept sensitive data should offer the possibility for controlled, verified access, and clearly document how to get access. When dealing with human data, additional requirements might come into play. For animal experimentation, ethical approval should also be appended.

## Follow best practices for licensing & responsibility

Digital services can simplify reuse of data and software. Understanding what can be done with downloaded resources is an important stage in the search for the right service.  We recommend service providers to clearly and prominently communicate all access and deposit conditions, and to state a license for downloaded data, software and derivatives.

To facilitate reuse, derived data and other downloaded resources should by default have clear standard licenses. Licensing is a way to standardize usage agreements and ensure that they are readable, clear, well thought through and don't result in unintended consequences. It is essential that a service uses appropriate licensing. We therefore recommend that all services use standard licenses wherever possible, and clear and readable data use agreements where licensing is not an option. Preferably the service is using one or several well known and easy to understand license models (e.g. Creative Commons, https://creativecommons.org/about/cclicenses/) at a clear and appropriate level of granularity. However, some types of data, including sensitive data, might come with conditions not easily covered by licensing. For such data, a readable yet sufficiently detailed Data Usage Agreement is required.

The service should be operated with information technology best practices, including communicating outages and changes, establishing a backup and archiving process, having excellent documentation and user support, performing security controls and updates, and allowing for privacy controls if needed.

Rights and responsibilities of both user and service should be articulated in a clear and transparent manner, with clear Terms of Use and an End User license or agreement. There should be policies for data preservation, authorship, and for scientific gateways with user accounts also a privacy policy and a code of conduct.

For digital resources to support reproducibility, it is important for the service to transparently communicate any changes, and the change history, in software used for data (pre)processing and analysis.

For both repositories and scientific gateways it is important that users are able to plan their work and schedule computational sessions there or elsewhere; we recommend that service providers clearly communicate planned outages in advance, via a dedicated channel (such as a dedicated mailing list) uncrowded by other communication

## Ensure accessibility and interoperability

Increasingly, big and complex neuroscience datasets need to be stored with or easily connected to computing resources. A dataset might even be distributed across different repositories, for example to satisfy the demands for sensitive data, or for studies consisting of different modalities or groups of subjects needing specialized repositories. Interoperability of repositories will be crucial; ease and homogeneity of access will be key, and metadata has a very important role here.

A service can make itself more accessible, interoperable and useful to its target community by using established community standards, for both data and metadata, and community vocabularies. In neuroscience, the BIDS (Brain Imaging Data Structure) format for neuroimaging data[11], and the NWB (Neurodata Without Borders) format for electrophysiology data[12] both have strongly facilitated data sharing and collaboration, and the NeuroML markup language[13] has made it possible to clearly describe, share and reuse computational neuroscience models.

Offering submission in a community format saves users from having to reformat all their data, makes metadata ingestion easier to support and to automate, and results in clear and consistent naming. Broadly available data in community formats will also lower barriers to the development of a surrounding ecosystem of software tools. We also recommend having methods reported in a structured format, a community relevant format if possible.

When community standards are not relevant or available, using an applicable general standardization framework is a preferable alternative over designing a new, custom format; this choice increases the likelihood of data and metadata being possible to transform into a future community standard format.

Programmatic or command-line access is vital for modern computational science to work seamlessly; and needs and demands on these types of access will likely change depending on community developments. We recommend that service providers offer an open, well documented API and/or a command-line interface (CLI) in several community relevant programming languages. Ideally, the API and/or CLI should also be open to community input.

Services should interact with community and other community relevant services to strive for interoperability, consistent access and authorization, and use of community vocabularies.

## Build capabilities for reproducibility, replicability, reuse

Data repositories and scientific gateways have the potential to contribute strongly with technical reproducibility and consistent data quality; ideally, they should make reproducibility as easy as possible. Machine readability and technical findability has to be considered.

The use of (machine readable) persistent identifiers (PID) is a core requisite for making research data accessible and fulfilling the FAIR principles. Services should assign PIDs to data descriptions, data and complimentary materials, assign PIDs in an appropriate metadata field when registering research data and ensure that descriptions of data always include a clear PID link to the data. We recommend that services use permanent identifiers at least for data (DOI) and software (DOI + SWHID[14]), ideally also for authors (ORCID) and associated research objects (RRIDs[15]). We also recommend that service providers register for an RRID that refers to their infrastructure.

Metadata are critically important to FAIR; they are the backbone of any dataset, and ongoing quality control of metadata is as important as the data. They are vital in ensuring that data can be correctly understood and effectively used and reused.

We recommend services to document and communicate their curation processes for data and metadata. Where possible, higher level curation which links to annotation and other published information material is preferable.

We recommend that methods are reported in a structured, community relevant format, (examples: STAR Methods, MDAR) and that metadata entry is made easy and automatically or semi-automatically verified. Ideally, methods are also published and citable (using platforms such as protocols.io).

We recommend that key software such as analysis algorithms is versioned and documented, and the versioning history communicated. Provenance for data, derived data and software should be documented and extractable.

We recommend that versioning of both content and authorship is transparently communicated and available for datasets, code, and analysis software.

Search and data discovery is an important step on the road to reuse. Providing easy data discovery and an overall user-friendly experience is a key service to all data repositories. We recommend services to interact with their community to identify and accommodate various data search behaviours, and to deliver search summaries that make it possible for researchers to judge relevance, accessibility and reusability of a data collection from the summary.

## Excel in documentation & user support

Documentation saves time, frustration and resources. The importance of good documentation cannot be overstated, ideally that documentation is also updated regularly and includes community input.  We recommend service providers to have extensive, clear and readable documentation.

Providing sufficient user support is an essential criterion.  Even with good documentation, it can take some time and effort for first time users to orient themselves. We recommend that all service providers have a FAQ with the most common user questions; ideally also a quick start guide. We further recommend that the service providers provide training materials specific to the service; ideally that they also provide or refer users to other relevant training on such issues as FAIR and reproducibility.

Supporting users is time and resource consuming, but can be made less so if the user community is provided with means to support themselves. For providers that do not want to set up their own support infrastructure, we suggest using an existing community forum that allows providing a dedicated channel or tag. Several neuroscience community tool projects already use the Neurostars forum (neurostars.org) this way. We recommend that service providers enable community users to support each other by setting up or utilizing mechanisms such as a user forum or mailing list.

## Be transparent in governance

Users are unlikely to rely on services they do not trust. In research infrastructures trust requires transparency at all levels, from governance to issue handling and communicating updates, outages and changes. We recommend that service providers document and clearly communicate the governance process, including issue reporting and resolving. Ideally, community users should have influence on governance. Funding sources, or other contributions of value, should be transparently communicated, and any conflicts of interest should be declared.

## Involve community in governance & decision making

For a service to be FAIR, its data and metadata need to meet domain relevant community standards, in order to increase their usability for their intended community. If a community has standards or best practices for data archiving and sharing, services should aim to implement and follow these standards.

Implementing community standards is a start at being community relevant, but it will be hard for services to achieve lasting broad community usefulness and impact without also having a mechanism for continuous community input and influence; therefore we recommend that all service providers include their intended and actual community in their setup and decision making process. At a basic level this can involve feature wishlists or feature upvoting, at a more advanced level also the possibility to nominate persons to participate in the steering committee or other governing body.

## Be transparent on sustainability - financial & technical

Research results are meant to last, and to be possible to revisit and reuse. Therefore, trust and long-term security are important factors in choosing a service for research activities and outputs.

The sustainability of a service, both financial and technical, is a key selection criteria. Sustainability support is often lacking in grant schemes [Khodiyar et al 2021]. Transparence on sustainability is important to allow researchers to make informed decisions about what services they use and invest their resources in.

Financially, we recommend that service providers transparently communicate current and past grants and other financing. Technically, we recommend that a service provider creates and provides transparent plans for archiving and backup, service closure and data and metadata preservation, and that they support sustainability by using open, established and maintainable technologies in their services.

The sustainability plan should address shutdown and archiving matters such as archiving or data preservation. The service should also state its data preservation policy.

We also recommend that the sustainability of the governing body itself is made clear, especially if it is not naturally renewed by elections.

## Conclusion

These recommendations and their associated criteria were developed with the intent to fit repositories as well as scientific gateways. They cover a range of important areas for users selecting digital services, including accessibility, licensing, community responsibility and the technical and financial sustainability of a service. Transparency and clear communication with users is a common denominator for many of the recommendations.

The recommendations were developed from a neuroscience perspective, but several of them are general and domain agnostic - they can apply to data repositories as well as software repositories and science gateways in any scientific field - because they deal primarily with how a service is run and governed.

As a research field, neuroscience has many different communities at very different stages of digital maturity. We hope that our high level recommendations will fill a bridging role and look forward to helping communities develop roadmaps towards adopting and implementing them.